\documentstyle{l-aa}
\input{psfig.txt}

\begin{document}

\thesaurus{11.09.1:
	   11.09.2:
	   11.09.4
	   }

\title{The filament of ionized gas in the outskirt of M87.
\thanks{Based on observations taken at INT and OHP. 
The INT is operated by the I.N.G. at the Spanish
Observatorio del Roque de Los Muchachos of the I.A.C. . The 
OHP is operated by the French CNRS.
}}


\author{G. Gavazzi\inst{1}
\and A. Boselli\inst{2}  
\and J. M.  V\' \i lchez \inst{3} 
\and J. Iglesias-Paramo\inst{4} 
\and C. Bonfanti\inst{1}}

\offprints{G. Gavazzi}

\institute{
Universit\`a degli Studi di Milano - Bicocca, P.zza dell'Ateneo Nuovo 1, 20126 Milano, Italy.
\and 
Laboratoire d'Astronomie Spatiale, Traverse du Siphon, F-13376 Marseille Cedex 12, France.
\and
Instituto de Astrof\' \i sica de Andaluc\' \i a, CSIC, Apdo.  3004, 18080, Granada,  Spain. 
\and
Instituto de Astrofisica de Canarias, 38200 La Laguna, Tenerife, Spain.
}

\date{Received..........; accepted..........}

\maketitle

\markboth{G. Gavazzi et al.: the outer M87 filament}{}

\begin{abstract}

We report on the filament of ionized gas found at the
NE periphery of M87, the second brightest elliptical galaxy in the Virgo cluster.
The object lies at 3.16 arcmin (15.7 kpc) projected distance of the nucleus
of M87, and it coincides in position with the Eastern radio lobe of M87.
Long slit spectroscopy confirmed the gaseus nature of this object,
showing $\rm H_{\beta}$, [OIII], [OI], [NII], $\rm H_{\alpha}$ and [SII] lines, 
without any underlying continuum.
The redshift of the object, $V = 1170 \pm 25 ~km~s^{-1}$,
agrees with the velocity of M87 itself ($V = 1280 ~km~s^{-1}$),
and with the bulk of the ionized material found near the center of M87.
The $\rm H_{\alpha}$ flux of this filament corresponds to $ 1 \%$ of the
ionized material in M87.
The radio-optical coincidence together with the spectral characteristics
suggest that the filament is ionized by a shock.

\keywords{Galaxies: individual: M87; interactions; ISM}
\end{abstract}

\section {Introduction}

The filamentary $\rm H_{\alpha}$ structure associated with M87 on a scale of
approximately 2 arcmin, discovered by Arp (1967),
was studied by Ford \& Butcher (1979) and more recently by 
Sparks et al. (1993). The morphology and kinematics of the ionized gas led
these authors to consider, among other possibilities, a scenario
where a gas-rich galaxy is being cannibalized by M87. High resolution HST observations
of the inner 1 arcmin were consistent with this picture and provided evidence
that the infalling material is feeding the central black-hole in M87 
(Harms et al. 1994; Ford et al. 1994).
We serendipitously included M87 in a narrow band image
of the center of the Virgo cluster aimed at studying the $\rm H_{\alpha}$ emission from VCC1313,
a dwarf galaxy near M87. We used a camera with a field of view of $34 \times 34$
arcmin which allowed us to include the lump of ionized material at the NE periphery of M87, 
which was reported by Baum et al. (1988) to coincide in position with the Eastern radio-lobe
of M87.
Baum \& Heckman (1989) discussed the statistical properties of the line-emission associated
with powerful radio galaxies and concluded that the mechanisms responsible
for the ionization in these objects is either photoionization by UV photons of nuclear
origin, or ionization by shocks.
To confirm kinematically the connection of the outer filament  
with M87, and to constrain observationally its ionization properties 
we took a low dispersion spectrum of the filament.   
The present paper reports on these findings.

\section{Observations}

We took observations of M87 through 
a narrow band filter centered at 6568~\AA ~(95 ~\AA~bandpass).
The image was obtained in the photometric night of
feb 21, 1999, in 1.2 arcsec seeing conditions,
with the Wide Field Camera at the prime focus of 
the Isaac Newton Telescope (INT) at La Palma. The camera relies on  
4 Chip Mosaic of thinned AR coated EEV 4K $\times$ 2K CCDs,
with pixels of 13.5 $\mu m$ x 13.5 $\mu m$ (0.33 arcsec x 0.33 arcsec),
giving a field of view of $34 \times 34$ arcmin.
The ON-band filter was selected 
to cover the redshifted $\rm H_{\alpha}$ and [NII] lines. The underlying
continuum was taken through the Johnson R filter. 
The integration time was 4500 sec (ON-band),
split in three shorter exposures to get rid of the
cosmic rays, each dithered by 18 arcsec to recover the
gap between the 4 chips. The red-continuum frame was exposed 100 sec 
to avoid saturation on the nucleus of M87.
The photometric calibration was obtained exposing the spectrophotometric star Feige 56
on each of the 4 CCD chips. 

We obtained low dispersion spectra of the object with the $\rm 1.93~m$ 
telescope of the Observatoire de Haute Provence (OHP),  
equipped with the Carelec spectrograph (Lemaitre et al. 1990) 
coupled with a 2K $\times$ 1K EEV CCD, giving a spatial scale of 0.6 arcsec per pixel.
The observations were carried out in the 
night of February 3, 2000 in approximately 2 arcsec seeing conditions
through a slit of 5 arcmin $\times$ 2.5 arcsec.
The selected grism gives a spectral resolution of 
900 ($\rm 133 ~\AA$/mm or 1.79 \AA/pix) and a
spectral coverage in the region $\rm 3200-7100~\AA$.
containing the redshifted $\rm H_{\beta}$ ($\rm \lambda~4861.3~\AA$),
[OIII] ($\rm \lambda~5006.8~\AA$),
$\rm H_{\alpha}$ ($\rm \lambda~6562.8~\AA$), 
the [NII] doublet ($\rm \lambda\lambda~6548.1, 6583.4~\AA$) 
and the [SII] doublet ($\rm \lambda\lambda~6717.0, 6731.3~\AA$).  
To ensure having most light from the target object
(invisible on the TV camera) in the slit 
we took spectra along two P.A. (29 and 314 degrees), with the slit aligned with some
reference stars.
The object was exposed 60 minutes and
a final spectrum was extracted over the wavelength range $\lambda$ $\lambda$ 4000-7000 \AA.
The calibration was obtained on the spectrophotometric star Feige 34.
Both images and spectrum were reduced using standard procedures within the IRAF
environment. 

\begin{figure}
\centerline {
\psfig{figure=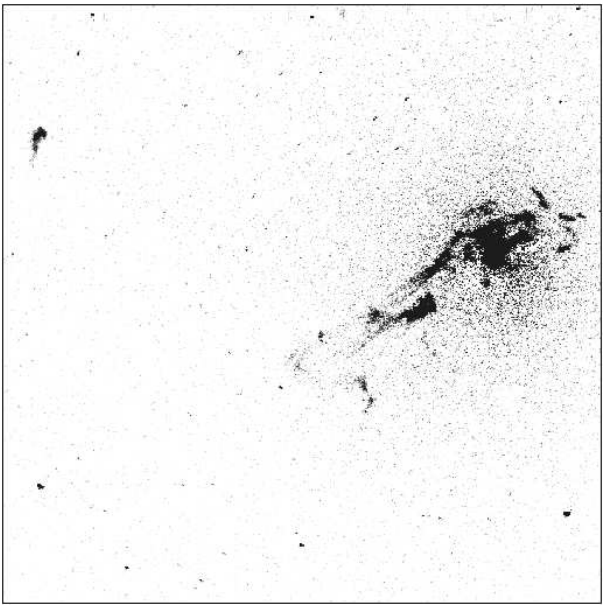,width=8cm,height=8cm}
}
\caption{The filamentary structure of M87 is visible to the right of this $\rm H_{\alpha}$ NET image ($4 \times 4$ arcmin field). The filament is visible at the upper left. North is up, East to the left.}
\label{fig.1}
\end{figure}

\section{Results}

Figure 1 gives the the continuum subtracted (NET) image of M87, showing the 
filamentary $\rm H_{\alpha}$ structure of M87. 
We obtain a total $\rm H_{\alpha}+[NII]$ flux of $1.18\times10^{-12}~ 
erg~cm^{-2}~s^{-1}$,
in very good agreement with Sparks et al. (1993)
At the NE periphery of M87 ($R.A.=12^h31^m02.14^s; Dec=12^o24'11.0"$; J2000)
we detect the presently discussed filament of $\rm H_{\alpha}$ emission
(the astrometry was performed relative two stars in HST Guide Star Catalogue within 2 arcmin
from the filament).
This corresponds to 3.16 arcmin (15.7 kpc) projected distance from the nucleus
of M87 (assuming a distance to M87 of 17 Mpc), thus the object lies inside
the stellar envelope of M87 whose isophotal radius at the $25^{th}
~mag~arcsec^{-2}$ level is 4.2 arcmin.

\begin{figure*}
\centerline {
\psfig{figure=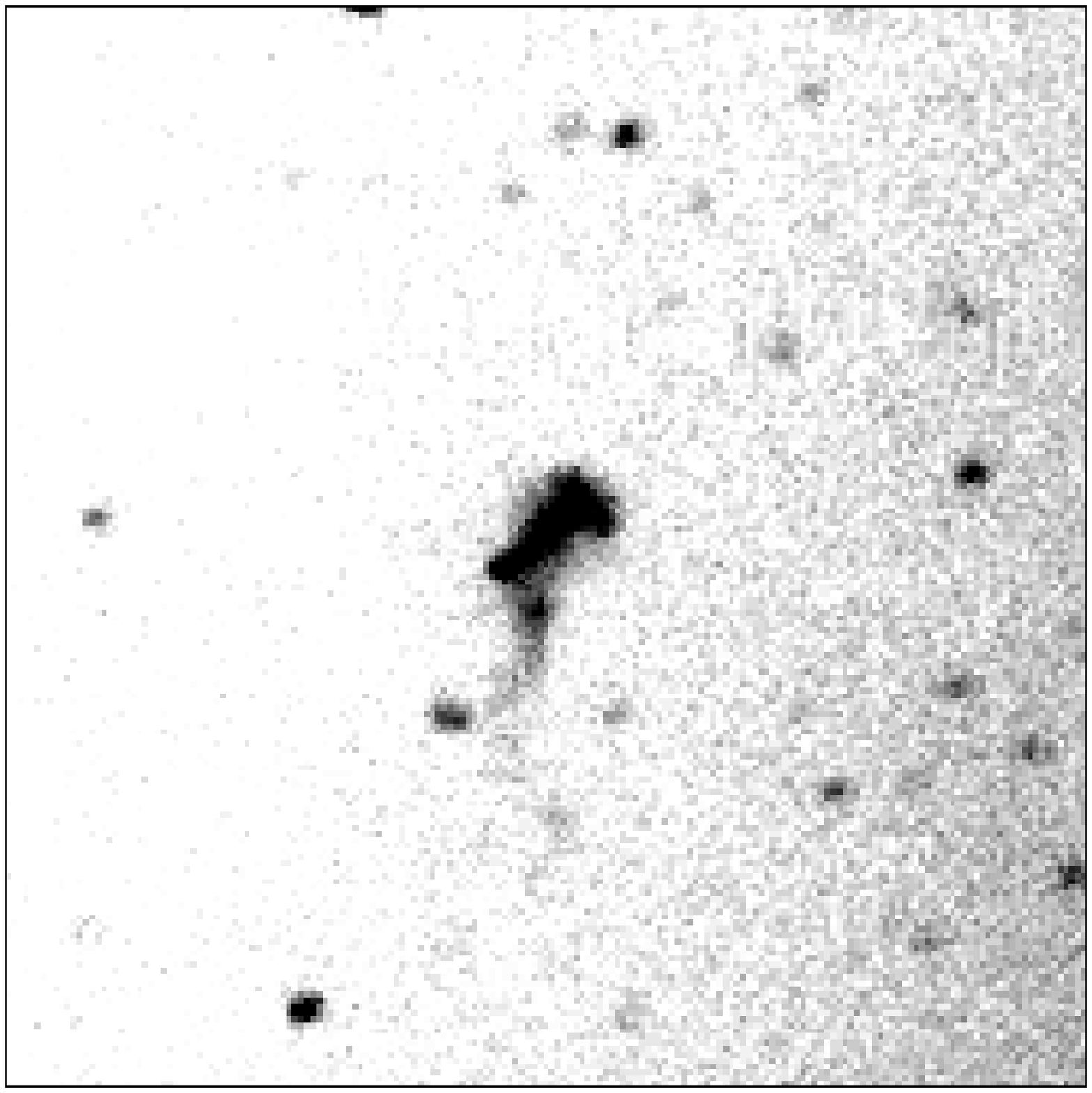,width=8cm,height=8cm}
\psfig{figure=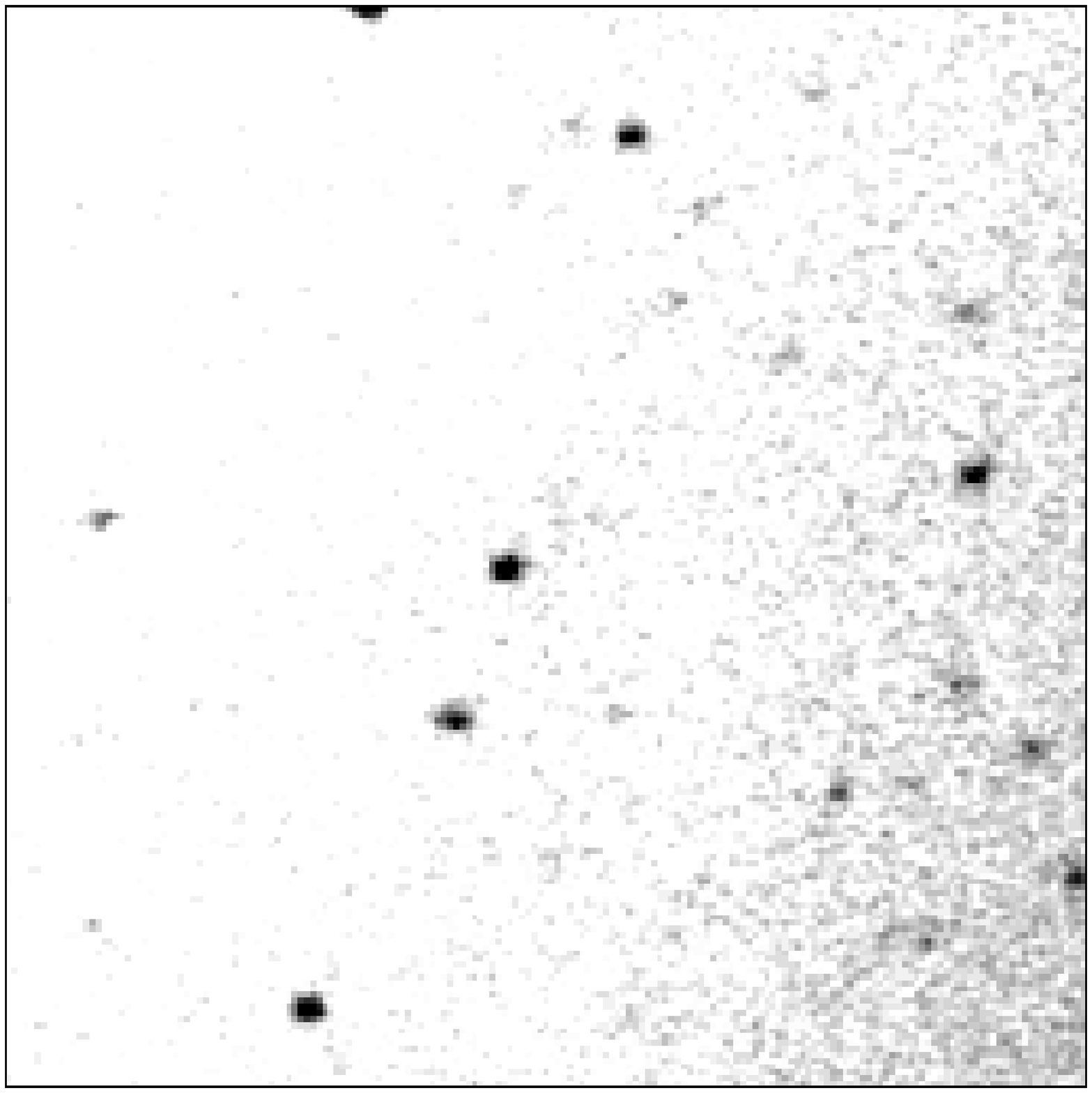,width=8cm,height=8cm}}
\caption{Enlargements of the $1 \times 1$ arcmin region containing the filament in the
$\rm H_{\alpha}$ ON-band (left; a) and in the red-continuum (right; b) frames. North is up, East to the left.}
\label{fig.2}
\end{figure*}

\begin{figure*}
\centerline {
\psfig{figure=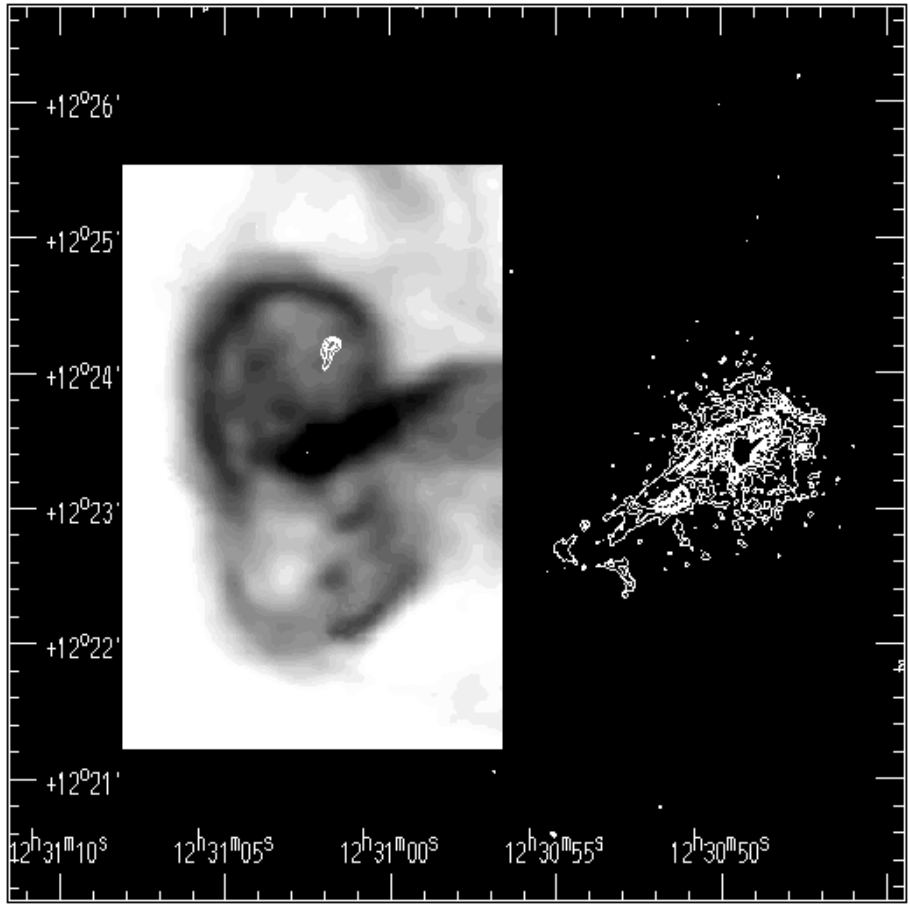,width=12cm,height=12cm}}
\caption{The $H\alpha$ net image (arbitrary contours) superposed onto the
East radio lobe of M87 (grey-scale, adapted from the 327 MHz map kindly provided by F. Owen). J2000 celestial coordinates are given.}
\label{fig.3}
\end{figure*}

Figure 2 gives enlargements of the region containing 
the object, which
is clearly detected in the ON-band image (Fig. 2a), while the total absence
of extended underlying continuum is evidenced by the lack of detection in the 
red-continuum frame (Fig. 2b). The central faint star is at $R.A.=12^h31^m02.14^s; Dec=12^o24'11.0"$; J2000 ($R.A.=12^h28^m30.34^s; Dec=12^o40'45"$; B1950). The object has a bright northern feature extending 6 arcsec to the N of this position, 
along with a faint southern tail (with 14 arcsec extension). The maximum extension is thus
20 arcsec N-S, and  8 arcsec in the E-W direction.
Altogether the filament has a "bow shape" appearance. 
The $\rm H_{\alpha}$+[NII] flux from the object is 
$10^{-14}~erg~cm^{-2}~s^{-1}$, corresponding to approximately 1\% of the
ionized material in M87.

We checked the positional corrispondence between the filament and the "large-scale"
radio source associated with M87. This is given in Fig. 3, where the 327 MHz radio map 
 kindly provided by F. Owen (with 5 arcsec resolution) is overlayed on top of our $\rm H_{\alpha}$ net image.
The filament is found within the eastern radio lobe of M87,
in contrast with Baum et al. (1988) who position the filament on the eastern-edge 
of the radio lobe, at approximately $R.A.=12^h28^m32.5^s$~(B1950), i.e. 30 arcsec
to the E of our position (see their Fig. 35). 
The exact location of the filament is inside the northern "arc" which constitue the
radio lobe, near to its N-W edge, abour 30 arcsec to the north of the tip of the radio beam. 

The red part of the spectrum of the filament is shown in Fig. 4.
We detect strong [NII] doublet bracketing the $\rm H_{\alpha}$ line,
the [SII] doublet and weak [OI]. More in the blue we detect weak $\rm H_{\beta}$ 
and [OIII]. No trace of continuum is present in the spectrum.
The line parameters are listed in Tab.1.
For each line (Col. 1) the observed wavelength (\AA) is given in Col. 2,
followed by the observed recessional velocity (not corrected to heliocentric)
in Col. 3, and by the intensity relative to $\rm H_{\alpha}$ (Col. 4)
(unexpectedly the ratio [NII] $\lambda~6548$ / [NII] $\lambda~6583$ differs from the canonical
0.3).
The average redshift of the object (excluding the discrepant
[OIII]) is $V=1170 \pm 25~km~s^{-1}$.
No significant line broadening is detected, as 
the line widths of [NII], $\rm H_{\alpha}$ and [SII] are found consistent with 6.2 \AA ~i.e.
with the width of the sky lines.

\begin{figure*}
\vskip -2 truecm
\hskip 3 truecm
\psfig{figure=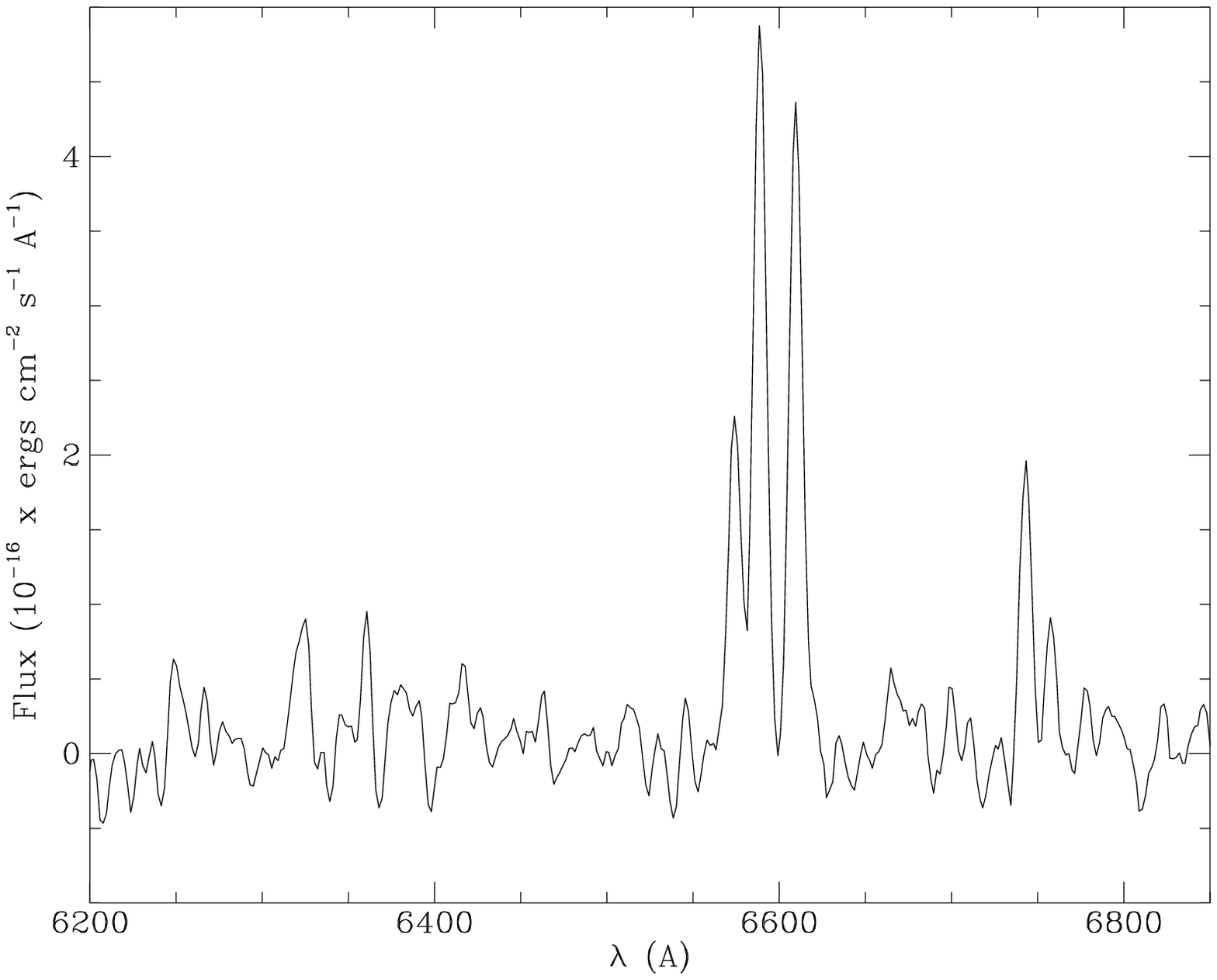,width=12cm,height=12cm}
\caption{The red part of the spectrum of the filament containing [OI], [NII], $H_{\alpha}$
and [SII].}
\label{fig.4}
\end{figure*}

\begin{table}
\caption{The line parameters}
\label{Tab1}
\[
\begin{array}{p{0.15\linewidth}cccc}
\hline
\noalign{\smallskip}
 (1)  & (2) & (3) & (4) \\
\noalign{\smallskip}
\hline
\noalign{\smallskip}
                & \lambda &  V   & flux \\

$H_{\beta}$     & 4879.65 & 1131 & 0.26 \\
$[OIII]$	& 5028.24 & 1285 & 0.22 \\
$[OI]$		& 6324.20 & 1152 & 0.18 \\
$[NII]$		& 6574.10 & 1191 & 0.45 \\
$H_{\alpha}$    & 6588.70 & 1184 & 1    \\
$[NII]$		& 6609.56 & 1192 & 0.90 \\
$[SII]$		& 6743.03 & 1163 & 0.34 \\
$[SII]$		& 6557.56 & 1170 & 0.19 \\
\noalign{\smallskip}
\hline
\end{array}
\]
\end{table}

\section{Discussion}

The most plausible interpretation of the observations presented in this
paper is that the filament of ionized material found in the outskirt of M87
is associated with the ionized gas found
near the nucleus of M87. In fact its redshift corresponds within 100 ~$km~s^{-1}$
to that of M87 and of the filamentary ionized gas.
Based on the estimate of Sparks et al. (1993) the ionized gas associated
with M87 amounts to only $10^{5-7} ~M_\odot$, thus the filament itself must 
contain a tiny fraction of gas $\sim 10^{3-5} ~M_\odot$. 
The plasma density derived from the $[SII] \lambda 6716 / [SII] \lambda 6731$
ratio is near to the low density limit of $10~cm^{-3}$, i.e. one or two 
orders of magnitude below ordinary HII regions (Osterbrock, 1989).
Besides the irrelevant mass and plasma density of the filament,
its strong [NII] intensity as well as the fluxes of [OI] and [SII] relative to H$\alpha$,
together with the [OIII] to H$\beta$ ratio are not typical of HII regions, nor
of Seyfert galaxies, but coincide with the zone of LINERS in diagnostic diagrams 
(e.g. Veilleux \& Osterbrock 1987). 
Moreover the intensity ratios observed in the filament differ markedly from those 
found in the central part of M87 and in other cooling-flow systems by Heckman et al. (1989).
The $[NII] \lambda 6584)/H\alpha$ and [SII]/H$\alpha$ are a factor of two lower than in the center
of M87, indicating a lower degree of ionization. 
Because of the irregular filamentary morphology of the source and the lack of extended underlying
continuum, photoionization by stellar sources -- hot young stars (Kim 1989),
post-AGB stars (Binette et al. 1994) -- seems ruled out. Similarly ionization by UV
photons originated in the active nucleus of M87 seems unlikely
because the filament is far from the nucleus.
Other mechanisms cannot be ruled out, such as for example X-ray heating by the ICM 
(Donahue \& Voit 1991), as suggested by the correspondence with the X-ray
elongation found by Harris et al. (2000). This class of models successfully explains 
the high H$\alpha$/H$\beta$ and [N{\sc ii}]/$H_{\alpha}$ ratios as
well as the $H_{\alpha}$ surface brightness of the filament
and was invoked by Heckman et al. (1989) and by Sparks et al. (1989)
as the likely explanation for the presence of ionized gas in cooling flow systems.
The ionization mechanism which provides the best consistency with the line diagnostics
of the filament is by intermediate velocity shocks (Dopita \& Sutherland 1995).
This is consistent with the positional correspondence found with the Eastern radio-lobe
of M87, which strongly argues in favour of the presence of shocks.
It is unclear, however, why the filament is confined to a small volume within the larger 
shocked region
and does not coincide exactly with any of the shock surfaces, nor with the tip of
the radio beam, as it would be expected if the gas was isotropically supplied by the IGM. 
It probably indicates that the ionization occours on a small lump of gas left over from
the gas-rich galaxy which was swollowed by M87.

\acknowledgements

This study was partly financed by the Spanish DGES (Direccion
General de Ensenanza Superior) (grant PB97-0158). 
The OHP observations have been supported by the french 
"Groupe de recherche Galaxies".

\end{document}